\documentclass[%
 reprint,
groupedaddress,
showpacs,
 amsmath,amssymb,
 aip,
]{revtex4-1}

\usepackage[final]{graphicx}
\usepackage{dcolumn}
\usepackage{bm}

\begin{document}


\title{Graphene based sensors: theoretical study}
\author{Karolina Z. Milowska}
 \email{Karolina.Milowska@physik.uni-muenchen.de}
 \affiliation{%
Photonics and Optoelectronics Group, Department of Physics and Center for NanoScience (CeNS), 
Ludwig-Maximilians-Universit{\"a}t M{\"u}nchen, Amalienstr. 54, 80799 Munich, Germany \\
 Nanosystems Initiative Munich (NIM), Schellingstr. 4, 80799 Munich, Germany
\\
}
\author{Jacek A. Majewski}%
 \email{Jacek.Majewski@fuw.edu.pl }
\affiliation{%
 Institute of Theoretical Physics, Faculty of Physics, University of Warsaw, ul. Ho\.za 69,
00681 Warszawa, Poland
\\
}%

\date{\today}

\begin{abstract}
Graphene, a 2-dimensional monolayer form of sp$^2$ hybridizated carbon atoms, is attracting increasing attention due to its unique and superior physicochemical properties. 
Covalently functionalized graphene layers, with their modifiable chemical functionality and usefull electrical properties, are excellent candidates for broad range of sensors, suitable for biomedical, optoelectronic  and environmental applications.
Here, we present extensive study of transport properties of sensors based on covalently functionalized graphene monolayer (GML) with graphene electrodes.  
The transmissions, density of states and current-voltage characterisctics supported by analysis of charge distribution of GML functionalized by -CH$_3$, -CH$_2$, -NH$_2$, -NH and -OH-CH$_3$, -CH$_2$, -NH$_2$, -NH and -OHGreen's function (NEGF). Further, we demonstrate how to control the device sensitivity by manipulating: (i) concentration, (ii) particular arrangement, and (iii) type of surface groups.  
We explain the underlying detection physical mechanisms. Comparisons of the theoretical results to available experimental data are provided and show good agreement.
\end{abstract}


\maketitle   

\section{Introduction}
Nowadays graphene, a 2-dimensional semimetalic monolayer, a form of sp$^2$ hybridizated carbon, attracts a lot of research activity owing to its  unique electronic, mechanical and thermal properties\cite{novoselov2007}. Graphene layers (GLs) are emerging as the very promising candidates for a new generation of electronic devices\cite{park2009,pacile2011,sengupta2011,georgakilas2012}.
From electronic  point of view, graphene has several properties  such as  extremely high mobility, high surface area-to-volume ratio, fast electron transfer rate, ballistic transport on submicrometer scale, the electron conductivity even higher than for copper at room temperature  or good biocompability\cite{luque2010,neto2009b,sarma2011,kuila2012,pramanik2011,basu2012,peres2010}, which can be beneficial in the design electrochemical sensors. 
However, pure graphene monolayer (GML) has zero energy band gap, which hinders direct application of graphene layer in digital electronics. The covalent functionalization of GLs is one of the possible solutions\cite{mao2013} to this problem.  It  opens the band gap and facilitates the usage of graphene  monolayers as sensors \cite{georgakilas2012}.
It was already reported that graphene based sensors  can detect even  individual molecules\cite{liu2011a,basu2012,schedin2007}. Because of biocompability of graphene, it is possible to detect glucose\cite{wang2010a,liu2010}, proteins\cite{ang2008}, DNA\cite{lu2009} or even bacteria\cite{mohanty2008,chang2012}. Smaller, faster and more sensitive sensors (i.e., graphene based) are nowadays searched for environmental applications\cite{shao2010,kehayias2013, li2012}. 

Proper sensing depends on many parameters, which demand a trade off between mechanical and electrical properties such as interface accessibility with molecular sensitivity and selectivity at room temperature or mechanical and electrical robustness.
At pristine graphene  surface, therefore, the direct adhesion of the sensed substances to the graphene is rather unprobable. Functionalization with small molecules can resolve this problem enhancing chemisorption of target molecules to the basal plane of graphene.
Moreover, it was shown\cite{erni2010} that adsorption of molecules, such as primary amines, methyl or hydroxyl groups, is highly probable when graphene membranes are unavoidably exposed to air and various organochemical solutions during preparation. 

Unfortunately, functionalization, especially covalent one,  can lead to many unwanted changes in morphology, both on local and global scale, elastic and electronic properties\cite{boukhvalov2009,boukhvalov2010b,milowska2011,milowska2012,milowska2013}, and also modify transport characteristics. To analyse this effect, we have performed studies of transport properties of  GMLs functionalized with simple organic molecules such, as -NH$_2$, -NH, -CH$_2$, -CH$_3$, and -OH, at low concentrations. 

\section{Theoretical methods}

The electronic coherent transport  has been studied employing non-equilibrium Green's function (NEGF) technique, within the Keldysh formalism, a rather standard procedure for treatment of transport in coherent regime\cite{book4}. The studied structures have been treated as two-probe systems with the central scattering region sandwiched between semi-infinite source (left) and drain (right) electrode regions (as shown in Fig.~\ref{fig:Fig1} (a)). The transmission coefficient of electrons (incident at energy E) through the central scattering region constituting the device (under the bias voltage V$_b$ equal to the electrochemical potential difference between the left and right electrodes, eV$_b$ = $\mu_L - \mu_R $)  T(E,V$_b$) has been calculated using the following expression:
\begin{equation}
T(E,V_b )=\hat{\Gamma}^{1/2}_{R} \hat{G^{r} }\hat{\Gamma}^{1/2}_{L},
\label{eq:e1}
\end{equation}
where $\hat{\Gamma_{R(L)}}$ are matrices that take into account the coupling of the central region to the right(left) electrode, and $\hat{G^r}$  is retarded Green's function of the system. The conductance can be expressed as follows \cite{book4}
\begin{equation}
G(E)=G_o Tr [\hat{\Gamma}_{R} \hat{G}^r \hat{\Gamma}_{L} \hat{G}^a ],
\label{eq:e2}
\end{equation}
where $G_o = 2\frac{e^2}{h} $ is the unit of quantum conductance and $\hat{G}^a$  is advanced Green's function.
The current flowing through the scattering region has been calculated according to Landauer-Buttiker formula, i.e., assuming the limit of small bias, which is justified for the range of the external voltages considered in the present study: 
\begin{equation}
I(V_b ) = \int\limits_{\mu _R }^{\mu _L } {T(E,V_b )dE} .
\label{eq:e3}
\end{equation}

We have used TranSIESTA \cite{brandbyge2002} package to calculate the transmission coefficients T(E, V$_b$) according to Eq.~\ref{eq:e1}, and then current-voltage I(V$_b$) characteristics due to Eq.~\ref{eq:e3}.
The source (left) and drain (right) electrodes contain 16 carbon atoms each. The central region, with electrode extensions, contains in each case 80 carbon atoms to which functionalized groups are attached.  For the geometry relaxations of central region without electrodes extensions (48 C atoms with adsorbates), we have used SIESTA package \cite{ordejon, soler, siesta3}.
The C atoms, adjacent electrode extensions were fixed to perfectly match with those regions meaning that functionalized structure is stressed (ca. 1.5 kbar).

The so-called PBE form of the generalized gradient approximation (GGA) \cite{Perdew1996} has been chosen for the exchange correlation density functional. Computations have been performed employing following parameters of the SIESTA package that determine numerical accuracy of the results: double-$\zeta$-plus-polarization basis, kinetic cut-off for real-space integrals of 500 Ry, the self-consistency mixing rate of 0.1, the convergence criterion for the density matrix of 10$^{-4}$,  maximum force tolerance equal to 0.001 eV/\AA, and  8x8x3 k-sampling in Monkhorst Pack scheme. 
For the transport calculations, the complex energy contour has been always set to value below the lowest energy in the energy spectrum of each system. The number of points along the arc part and on the line of the contour  have been set to 16 and 10, respectively, whereas number of Fermi poles to 16. For the current calculations, we have used default value of small finite complex part of the real energy contour (10$^{-6}$ Ry) and increased number of points on the close-to-real axis part of the contour in the voltage bias window to 10. For accuracy of transmission spectra, the energy window has been in the range of (-3,3) eV, the number of points for the computation of the transmission function has been quite high (500) and the number of eigenvalues of the transmission matrix has been set to 3.
All calculation have been performed  with spin polarization.

\section{Results and discussion}

\begin{figure} [htbp]
	\centering
	\includegraphics[width=0.50\textwidth]{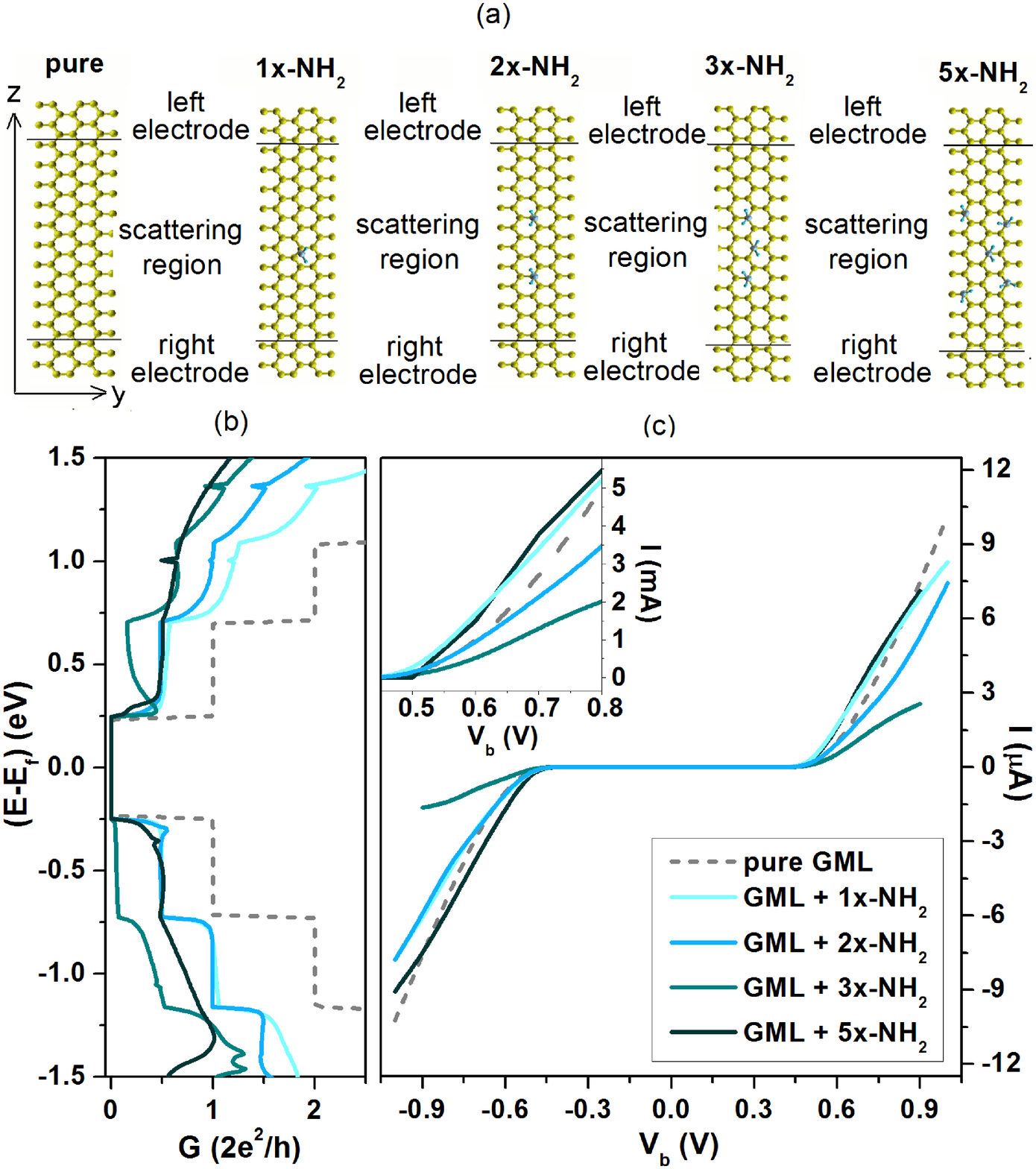}
	\caption{\label{fig:Fig1} 	(a) The schematic view of studied functionalized systems: pure graphene monolayer, graphene functionalized with one, two, three, and five -NH$_2$ groups. Electrodes (containing 16 C atoms each) and scattering regions (containing 80 C atoms) are also depicted by black lines.  
Periodic boundary conditions are conserved in perpendicular directions to the z axis. In x axis direction, separation between layers is large enough (80 \AA) to eliminate any interactions. The voltage is applied to the electrodes. 
	(b) The zero-bias transmission spectra for pristine (dashed line) and   functionalized GML (continuous lines) as a function of electron energy relative to the Fermi level. (c) I(V$_b$) characteristics of all systems for a bias voltage (V$_b$) from -1.0 to 1.0 V. Inset contains magnified region from 0.45 to 0.8 V.}
\end{figure}

Motivated by experimental data \cite{baraket2012}, we have started our studies of transport properties of functionalized GMLs with primary amines.
The schematic view of studied functionalized systems: pure graphene monolayer, graphene functionalized with one, two, three and five -NH$_2$ group are shown in Fig.~\ref{fig:Fig1} (a).

\begin{table}
\caption{
Structural properties of all studied systems: adsorption energy, $E_{ads/N}$,  and strain energy, $E_{strain/N}$. $E_{ads/N}$ has been calculated as $E_{ads/N}  = \frac{1}{N} \left( E_{GML + groups}  - (E_{GML}  + N \cdot E_{group} ) + E_{cc} \right) $, with $E_{cc}=E_{GML + \cdot ghost}  - E_{GML}  + E_{ghost + groups} - N \cdot E_{group} $ where $E_{GML + groups}$,  $E_{GML}$ and $E_{group}$ are total energy of functionalized system, pure GML and adsorbates, respectively. $E_{cc}$ is basis set superposition error correction. $E_{ghost + groups}$ and $E_{GML + ghost}$ are Kohn-Sham energies of the functionalized system where the adsorbates or GML are replaced  by their ghosts \cite{soler} and all atoms are fixed on optimized positions. $E_{strain/N}$ is basically the difference in total energies between functionalized system with removed adsorbates and pure relaxed GML divided by number of attachments.
System depicted by ${AA}$ has two -NH$_2$ groups on the same graphene sublatice, whereas system ${AB}$ has those attachments on different graphene sublatices. No. in first column indicates a number of adsorbates on GML.
}
\begin{tabular}{|l|c|c|c|}
\hline
 No. & Type & {$E_{ads/N}$ (eV)} & {$E_{strain/N}$ (eV)} \\
 \hline 
\hline
1  & -NH$_2$ & -0.314 & 1.215 \\
2  & -NH$_2$$^{AA}$ & -0.429 & 1.212 \\
2  & -NH$_2$$^{AB}$ & -0.442 &  1.155  \\
3  & -NH$_2$ & -0.537 & 1.192 \\
5  & -NH$_2$ & -0.412 & 1.303  \\
\hline
2  & -CH$_3$ & -0.779 & 1.132 \\
\hline
2  & -OH & -1.498 & 1.069 \\
\hline
2  & -NH & -1.509  & 1.064 \\
\hline
2  & -CH$_2$ & -1.893 & 1.283 \\
\hline
\end{tabular}
\label{tab:Tab1}
\end{table}

Chemisorption of -NH$_2$ groups leads to breaking one of $\pi$-bonds and  sp$^2$ to sp$^3$ rehybridization of C-C bonds. Unpaired electron, which forms this broken bond remains at the neighbouring C atom. The chemisorption of the next group on different sublattice is more energetically favourable than on the same sublattice
(i.e. for -NH$_2$ group the difference in E$_{ads/N}$ is equal to 0.033 eV per attachment, see Tab.~\ref{tab:Tab1}).
 Similar observation concerning hydrogen functionalization of graphene was presented by Boukhvalov and Katsnelson  in Ref.\cite{boukhvalov2009}. Therefore, it is more probable that in an experiment a reduction of conductance with growing density of amines will be observed rather than conservation of conductance at pristine graphene level.

Garcia-Lastra \textit{et al.}\cite{lastra2008} showed that there is a systematic dependence of transmission on the position of the adsorbed molecules to the lateral surface of metallic nanotubes.
The behaviour of the conductance $G$ as depicted in Fig.~\ref{fig:Fig1} (b) is characteristic for metallic carbon nanotubes with single adsorbed molecule placed on top of carbon atom, or more adsorbed molecules placed on equivalent lattice sites (say $A$) with certain positions determined by the vectors $\vec{R}=n \cdot \vec{a} + m \cdot \vec{b}$, with $ n - m = 3p$, where $m$, $n$, and $p$ are integers, and $\vec{a}$ and $\vec{b}$ are two primitive translations of the graphene lattice.  
This behaviour can be understood in terms of coinciding K, K' and $\Gamma$, the high symmetry points,  in graphene systems where groups  are adsorbed at sites $A$, being multiple of 3 \cite{lastra2010}. 
In the case studied in this paper, the two -NH$_2$ molecules were placed at graphene lattice sites with $(n,m)$ equal to $(0,0)$ and $(3,0)$ at so-called top positions, therefore, fulfilling the rule\cite{lastra2008} and indicating that graphene layer can be treated as metallic carbon nanotube of infinite radius. 
In the case of  three adsorbates,  one of them was placed on different sublattice (say $B$) leading to partial closing of the conductance channel.
However, for the case of five adsorbed amines, where molecules are placed at different graphene  sublattices and only two of adsorbates are fulfilling the rule, where three others not, the behaviour seems to be different. The conduction around Fermi level is bigger than for three groups and is almost as big as for the case of one and two adsorbates. But it should be noticed that this situation changes according to Garcia-Lastra rule outside the range  (-1.3,1.3) eV, where Fermi energy is set to 0.

No less interesting is an analysis of dependence of current-voltage characteristics on concentrations of -NH$_2$ groups, depicted in Fig.~\ref{fig:Fig1} (c).
The critical bias voltage for the non-zero current is the lowest for the system with one amine and the highest for five amines covalently bound (see inset in Fig.~\ref{fig:Fig1} (c)). 
For the bias voltage in the range of 0.6 - 0.9 V, the current for GML functionalized with one and five amines is larger than for not functionalized one, whereas for GML functionalized with two and three amines is smaller. For bigger bias voltage, one can observe that current of all functionalized systems is smaller than for pure graphene. However, as one can see, the I(V$_b$) characteristics of functionalized GML are not symmetric with respect to the point (0,0). 
For negative bias voltage, the current flowing through the system with five amines attached is significantly bigger than for a system with smaller  concentration of adsorbates.
There are two possible explanations: (i) the stress induced by adsorption of amines to graphene lattice and (ii) the interactions between groups.  The covalent bonding between GML and adsorbates leads to local (rehybridization of the C-C bonds, out-of-plane distortions) and global (elongation of the lattice constant $a$) modifications of graphene lattice and those effects are more pronounced for higher densities of  adsorbates.
The nonlinear current response  of graphene nanoribbons to bias voltage for higher strains was also reported by  Topsakal \textit{et al.} \cite{topsakal2010a} and Wu \textit{et al.}\cite{wu2013}. This decrease and increase of the current,  as shown in Fig.~\ref{fig:Fig1} (c),  is directly related to the changes in morphology of the considered systems under tension.  
On the other hand, the interactions between groups start to be important with increasing density of adsorbates on basal plane of graphene. The shape of the transmission spectra for five amines is significantly changed in comparison to smaller concentration of -NH$_2$ groups - it is much more smeared from step-like transmission of pure graphene.
The conduction dip at 1.364 eV, which is clearly seen for one, two and three -NH$_2$ groups covalently bound to GML, disappears for five adsorbates.
 Wu \textit{et al. }\cite{wu2013} showed that electronic properties of graphene nanoribbons are strongly affected by  presence of tensile strain, which can be easily related with changes in transmission spectra of these systems. 

However, as one can see the observed changes in  the I(V) characteristics (Fig.~\ref{fig:Fig1}(c)) are rather small in comparison to pure graphene monolayer. This is consistent with the experimental results of Baraket\cite{baraket2012}. The increasing concentration of primary amines covalently bound to graphene monolayer causes an increase in the chemical reactivity of the surface, while the electrical conductivity is decreased.  However, even highly aminated graphene, up to 20$\%$, is conductive enough to be used for DNA detection as bio-attachment platform in a biologically active field-effect transistors. Therefore, it is plausible that these functionalized structures can be commercially used as biosensors \cite{georgakilas2012,basu2012,shao2010}.

 \begin{figure} [htbp]
\centering
\includegraphics[width=0.50\textwidth]{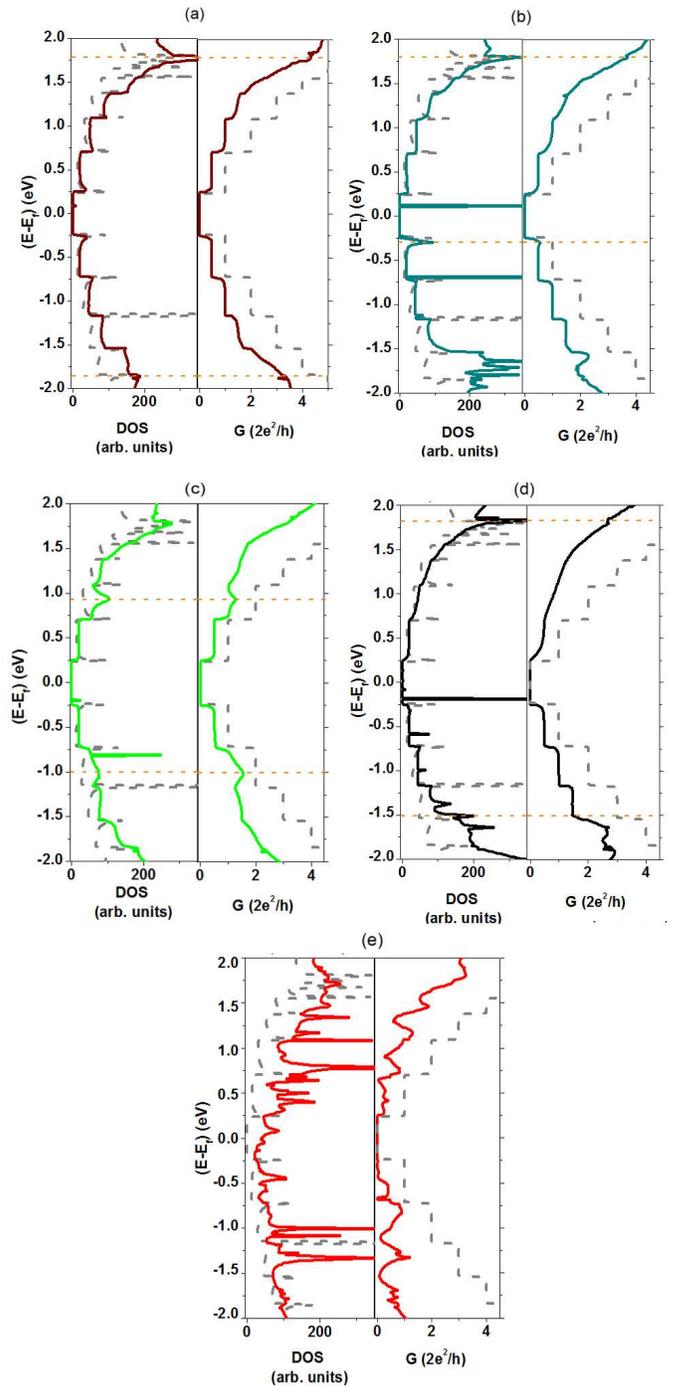}
\caption{\label{fig:Fig2} The density of states and the zero-bias transmission spectrum for GML functionalized  with (a) -CH$_3$, (b) -NH$_2$, (c) -NH, (d) -OH, and (e) -CH$_2$ fragments. The resonant/anti-resonant behaviours are indicated by orange lines.
The results for pure graphene are depicted by dashed gray lines in each graph.
}
\end{figure}

Each C atom from graphene lattice is capable of interaction with a single molecule and can be considered  as a possible target for gas or vapour species resulting in ultrasensitive response \cite{novoselov2007,basu2012}. It was shown that the adsorption of gas molecule leads to the local change in the carrier concentration inducing a doping of the delocalized 2D electron gas, which can be experimentally measured in transistor like devices \cite{pearce2011,ohno2010}.
Therefore, we have decided to test the possibility of detecting and distinguishing between very small concentrations of different
simple organic molecules.
In Fig.~\ref{fig:Fig2} we present DOS and transmission spectra of  GML functionalized  with five different types of adsorbates at the same concentration (two adsorbates): -CH$_3$,  -NH$_2$,  -NH,  -OH, and  -CH$_2$ fragments. The band gap is clearly seen in all systems, because only 48 C atoms of scattering region were optimized, in such way, that C atoms adjacent electrode extensions have to perfectly match with those regions.
Therefore, only at short range the graphene lattice can be affected by presence of the adsorbates and the structure is stressed (ca. 1.5 kbar). Impurity bands originating from attached groups are moved further from Fermi Energy  and do not cross the Fermi level in the case of hydroxyl groups.
 However, even for those systems, it is possible to distinguish between systems with different adsorbates (Fig.~\ref{fig:Fig2}) and its concentration (Fig.~\ref{fig:Fig1} (b)) from transmission spectra.

The -CH$_3$ and -NH$_2$ groups, as well as -NH radicals induce resonant transmission, whereas -OH anti-resonant transmission. 
The conductance peaks correspond with DOS peaks at -1.858 eV and 1.786 eV  for -CH$_3$ (see Fig.~\ref{fig:Fig2} (a)), at -0.295 eV and 1.798 eV for -NH$_2$ (see Fig.~\ref{fig:Fig2} (b)), and at -1.004  eV  and 0.932 eV for -NH radical (see Fig.~\ref{fig:Fig2} (c)). The adsorption of  -OH groups leads to  two small dips at -1.509 eV,  and 1.823 eV corresponding with DOS peaks (see Fig.~\ref{fig:Fig2} (d)).
These unique dips/peaks can be related with the type of adsorbates, because they are induced by quasi bond states in the structures. The  sp$^2$ $\to$ sp$^3$ rehybridization  caused by the covalent bonding of the adsorbates to the basal plane of GML is responsible for this picture. 
Let us remark that the conductance of graphene functionalized with amines drops much slower with concentration of dopants than in the case of  hydroxyl groups. It is mostly due to the fact that nitrogen acts as donor and has smaller electronegativity than oxygen. 
Transmission spectra of GML functionalized with -OH groups (Fig.~\ref{fig:Fig2} (d)) is much more smeared and asymmetrical around Fermi energy in comparison to pristine graphene and graphene functionalized with methyl group or amines. Hydroxyl group attracts electron cloud more efficiently than other considered adsorbates.
However, it is the case of graphene functionalized  with carbene that has a very different transmision spectra in comparison to pure graphene and to other considered attachments.
The transmission coefficients at 0.4 eV are equal to 0.329 0.393, 0.483, 0.488, 0.508 for -CH$_2$, -OH, -CH$_3$, -NH$_2$, -NH respectively.
Adsorbates break the hexagonal symmetry of GML and create scattering centres. In other words, adsorbates act as a carrier trap reducing carrier density.

We have shown that different adsorbates lead to different energy of resonant and anti-resonant transmission. This effect was also shown by Wei \textit{et al.} \cite{wei2012a} in the case of nitrogen-vacancy nanoribbons adsorbing NH$_2$, CO and N$_2$ molecules. 

\begin{figure} [htbp]
\centering
\includegraphics[width=0.50\textwidth]{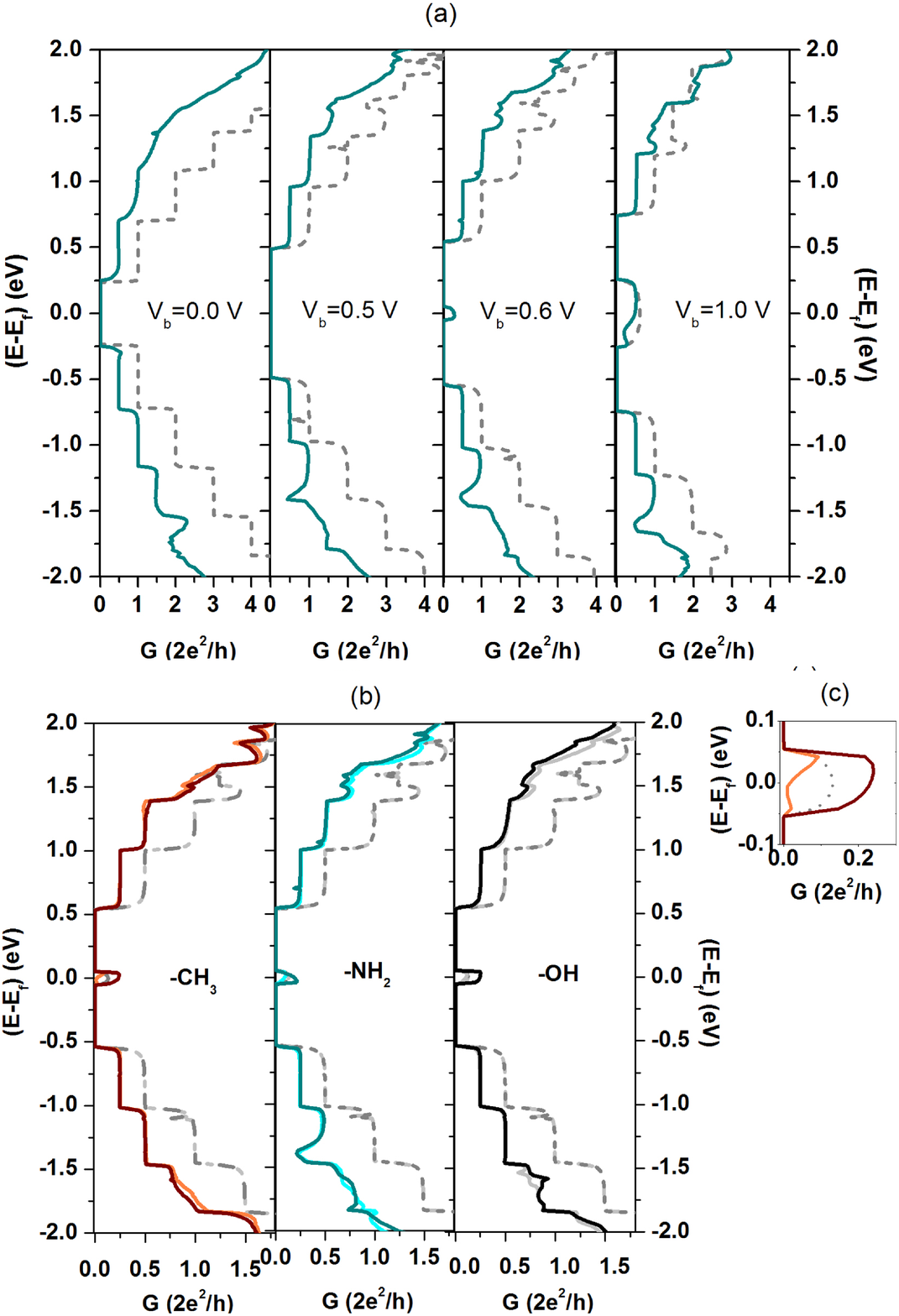}
\caption{\label{fig:Fig4} (a) The transmission spectra for  V$_b$ = 0.0 V, 0.5 V, 0.6 V, and 1.0 V for GML functionalized  with -NH$_2$. (b) The spin-resolved transmission spectra under V$_b$=0.6 V for systems functionalized with -CH$_3$, -NH$_2$, and -OH fragments; darker lines for spin-up electrons and lighter lines for spin-down electrones. (c) Magnified peak in transmission under V$_b$=0.6 V at Fermi energy for system functionalized with -CH$_3$.
The transmission spectrum for pure graphene (dashed gray lines) is also given for comparison.}
\end{figure}

In Fig.~\ref{fig:Fig4} we have presented transmission spectra for non-zero bias voltage.
Because of stress induced by fixing C atoms to perfectly match with those from electrode extensions, all studied systems have non-zero band gaps. 
 The conductance gap broadens near the Fermi level under applied finite bias voltage V$_b$=0.5 V in comparison to zero-bias voltage case. After exceeding threshold voltage, transmission spectra change. For applied V$_b$=0.6 V, one can observe an appearance of a peak at Fermi level (see (Fig.~\ref{fig:Fig4} (a))) due to strong group and electrode coupling. For V$_b$=1.0 V, this peak is broadened, its amplitude is increased to almost one and becomes asymmetrical. 

The spin-resolved transmissions spectra under V$_b$=0.6 V for all systems containing attachments with odd  number of electrons (-OH, -NH$_2$ and -CH$_3$) are shown in Fig.~\ref{fig:Fig4} (b). Both, spin-up (darker curves) and spin-down (lighter curves) transmission spectra differ. Close to Fermi level conducting channel for one spin population is suppressed more than for other one. The most striking difference can be observed for -CH$_3$ functionalized system, where electrons  with spin-down will be almost fully filtered, whereas electrons with spin-up will be transmitted (see Fig.~\ref{fig:Fig4} (c)). 
In systems containing -OH,-NH$_2$ and -CH$_3$ groups the degeneracy is lifted under the influence of the functional group to the whole system. The functional groups significantly change charge distribution on graphene layer by dipolar field. Therefore, this influence of attachments on graphene layer can be understood by Stark effect.

Upon application of finite voltage, two spin components carry electric current through device in different extent. For V$_b$=0.6 V, the differences in current between two spin populations are equal to 0.307 $\mu$A, 0.528 $\mu$A, and 0.656 $\mu$A for system containing two -NH$_2$, -OH and -CH$_3$ groups, respectively. The analysis of the conductance and current reveals  that those systems  can be  used as supramolecular spin-filtering devices within some specific energy windows.  
Next step is to analyse changes in current-voltage characteristics induced by different adsorbates.

  \begin{figure*} [htbp]
	\centering
\includegraphics[width=0.90\textwidth]{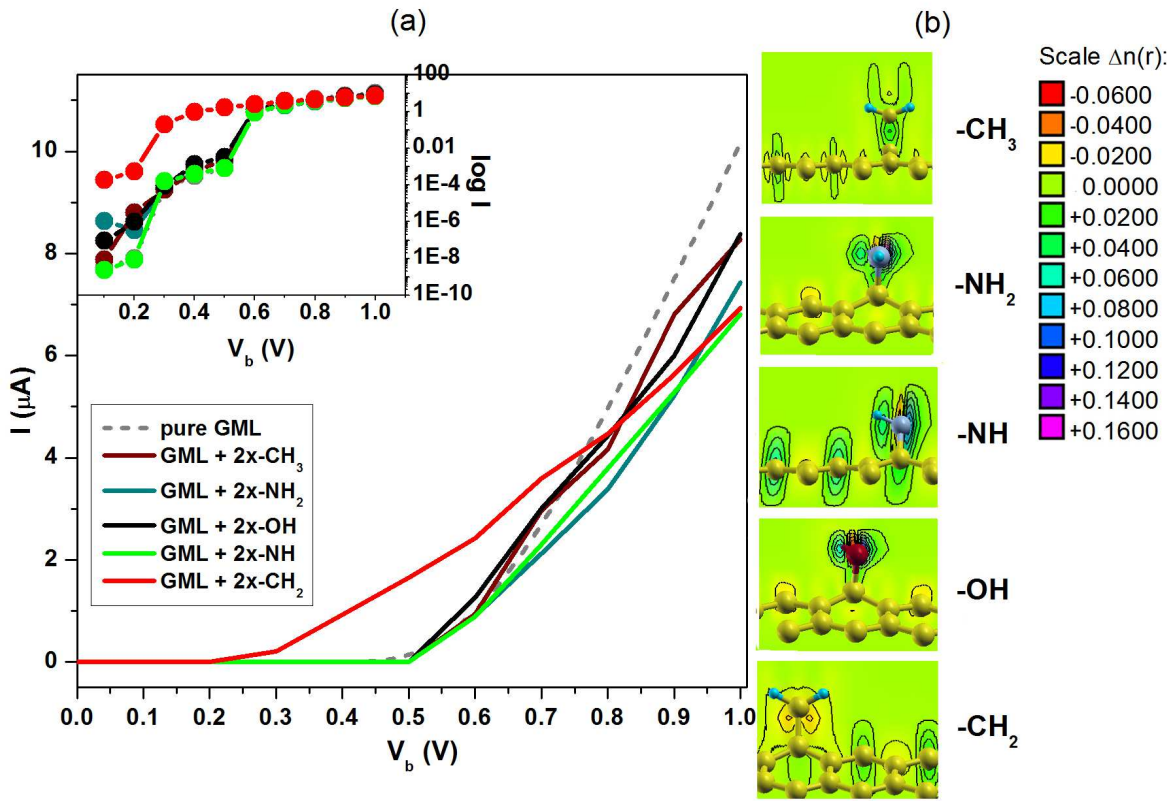}
\caption{\label{fig:Fig3} (a) The current-voltage characteristics of pure and functionalized GML for a bias voltage (V$_b$) from 0 to 1.0 V. Inset: logarithimc scale. (b) The difference of the valence pseudocharge density and the superposition of the atomic valence pseudocharge densities for all studied systems  plotted in the plane perpendicular to graphene layer, together with the ball-and-stick model of the functionalized systems. }
\end{figure*}

The calculated current-voltage characteristics (calculated using Eq.~\ref{eq:e3}) for GMLs functionalized with -NH$_2$, -NH, -CH$_3$, -CH$_2$ and -OH compared to the characteristics of the pristine GML (grey dashed line), are depicted in Fig.~\ref{fig:Fig3} (b). The critical voltage  for the non-zero current is the lowest  for GML functionalized with -CH$_2$ radicals (0.2 V). Those adsorbates cause the most significant changes in current-voltage characteristics - the slope of non-zero part for -CH$_2$ is  2.45 times smaller than for pristine graphene monolayer.
The I(V$_b$) characteristics for other adsorbates and pure graphene are similar; current gradients for GML functionalized with -CH$_3$, -OH, -NH$_2$, and -NH are   1.27-1.58 times  smaller than pure GML.

To better explain physical mechanism governing these effects, we have depicted in Fig.~\ref{fig:Fig3} (b) distribution of the changes of valence charge around adsorbates (side view) with the functionalized graphene lattice superimposed on. All studied types of adsorbates are strongly chemisorbed to the basal plane of graphene 
(see adsorption and strain energies in Tab.~\ref{tab:Tab1}). 
Depending on the electronic configuration, the fragments form either single chemical bond to the C atom from the GL basal plane (-NH$_2$, -CH$_3$,  and -OH), being in the so-called top position, or double bond (as -CH$_2$ and -NH) with fragment's C atom placed in the bridge positions. 
Incorporation of atoms with different level of electronegativity leads to different changes in local charge density. Therefore, functionalization with groups containing carbon  slightly changes the situation in comparison to the pure graphene. Methyl group induces small positive valence electronic charge in the middle of the bond between basal plane and group, whereas carbene causes larger changes. One can clearly see areas of additional valence electronic charge more shifted to the adsorbate than to basal plane of graphene. For -CH$_2$, there are also bigger changes in C-C bond in basal plane of graphene than for -CH$_3$ group. Covalent functionalization with amines, -NH$_2$ and -NH, leads mostly to generated excess charge  at N atom. Again, the radical introduces more changes to the basal plane. In the case of hydroxyl group, the oxygen atoms introduce comparable positive difference of the valence pseudocharge density and the superposition of the atomic valence pseudocharge densities. However, the charge distribution disturbance caused by -CH$_2$ radicals is the biggest one from studied examples of adsorbates. It is smeared over C atoms to which the radical is not directly bound (see the last DRHO of -CH$_2$ functionalized graphene in Fig.~\ref{fig:Fig3} (b)).

Despite changes caused by functionalization it should be emphasized that the functionalized systems conduct comparable  currents  to pure system at the same bias voltage applied.
At $V_b$=1.0 V, the currents are equal to 10.160 $\mu$A, 8.376 $\mu$A, 8.274 $\mu$A, 7.434 $\mu$A,  6.937 $\mu$A, and 6.808  $\mu$A  for pure GML, GML functionalized with -OH,  -CH$_3$, -NH$_2$,  -CH$_2$, and   -NH  fragments, respectively. 
Covalent functionalization of GMLs does not ruin current flow through the system, as it is observed in the case of single walled carbon nanotubes (SWNT)\cite{bouilly2011}. 
Similar changes in current-voltage characteristics for armchair nanoribbons functionalized with -CH$_3$ and -OH groups were presented in Ref.~\cite{tsuyuki2012}. It is worth to mention that our I(V$_b$) characteristics of functionalized GML look similar to experimental ones of reduced graphene oxide and graphene oxide\cite{singh2011}.
  
Generally, covalent functionalization, which is needed for sensing purposes, leads to a decrease of electrical conductivity.
 For fabrication of graphene based devices, such as sensors, the balance between conductivity and enhanced chemical reactivity should be carefully investigated.
Our studies show that  each kind of adsorbate and its concentration can be detected by its own unique resonant or anti-resonant peaks.  
Sensing substances in GML-based devices can be easily monitored through two-probe transport experiments. The GML are better candidates for building sensors than carbon nanotubes, where single walled nanotubes are not sufficient and usage of at least double walled nanotubes is required for proper operation of such devices\cite{bouilly2011,baraket2012}.

\section{\label{sec:level4}Conclusions}

We have shown that transport properties are affected by covalent functionalization.
The transmission and I(V$_b$) characteristics differ between various types of attachments. The resonant/anti-resonant behaviours induced by quasi-bond states in the structures can be observed. Carbene  introduces the most dramatic changes in transmission spectrum and current-voltage characteriscics, whereas  methyl groups cause the smallest change.  This effect allows one to distinguish between type and  concentration of adsorbates with extremely high sensitivity. It should be emphasized that changes induced by functionalization are not dramatic in comparison with pristine GML. Therefore, functionalized graphene monolayers could be, in contrary to functionalized SWNTs, used as a channel in FET based sensors.

\section{Acknowledgement}
This work has been supported by the European Founds for Regional Development within the SICMAT Project (Contact No. UDA-POIG.01.03.01-14-155/09). We acknowledge also support of the PL-Grid Infrastructure and of Interdisciplinary Centre for Mathematical and Computational Modeling, University of Warsaw (Grant No. G47-5). K. Z. Milowska acknowledges funding from the European Commission through the FP7-NMP programme (project UNION, grant No. 310250). 

\bibliography{biblio}

\end{document}